\documentclass[11pt,a4paper]{article}
\usepackage[T1]{fontenc}
\usepackage[utf8]{inputenc}
\usepackage{lmodern,microtype}
\usepackage[margin=2.5cm]{geometry}
\usepackage{amsmath,amssymb,booktabs,enumitem,graphicx,placeins,array}
\usepackage[numbers,sort&compress]{natbib}
\usepackage{xcolor,xurl,hyperref}
\graphicspath{{figures/}}
\hypersetup{colorlinks=true,linkcolor=blue!50!black,citecolor=blue!50!black,urlcolor=blue!60!black,pdftitle={Fabric retrofit scenario analysis under uncertainty for an urban housing stock using machine learning}}
\newcommand{\SAP}{\mathrm{SAP}}
\newcommand{\EI}{\mathrm{EI}}
\newcommand{\clip}{\operatorname{clip}_{[1,100]}}
\newcommand{\ind}{\mathbf{1}}
\newcolumntype{L}[1]{>{\raggedright\arraybackslash}p{#1}}
\title{\fontsize{18}{22}\selectfont Fabric retrofit scenario analysis under uncertainty\\ for an urban housing stock using machine learning}
\author{%
\normalsize
Yunfei Bai\textsuperscript{1},
Aaron Tesfa Tsion\textsuperscript{1},
Ra\'ul Rosales\textsuperscript{1},
Barbara Shollock\textsuperscript{1},
Wei He\textsuperscript{1}\thanks{Corresponding author: \href{mailto:wei.4.he@kcl.ac.uk}{wei.4.he@kcl.ac.uk}}\\[0.5em]
\normalsize\textsuperscript{1}Department of Engineering, King's College London, London, UK%
}
\date{}
\begin{document}
\maketitle
\begin{abstract}
Fabric retrofit is central to decarbonising housing, and planning it depends on expected savings and their uncertainty.
This study asks which upgrade saves most per dwelling and across a housing portfolio, how uncertain the savings are, and whether upgrade order matters.
We propose an uncertainty-aware framework that converts predicted changes in energy performance certificate scores after wall, roof and window upgrades into standardised energy-cost savings and emission reductions.
Six machine-learning models predict the 1--100 Standard Assessment Procedure (SAP) energy-cost and environmental impact (EI) emission scores from 17 certificate descriptors.
Each model's errors on similar dwellings give calibrated prediction intervals.
A global-best strategy applies the most accurate model throughout; a sample-specific strategy chooses each dwelling's model using these local errors.
The analysis covers 103,414 officially registered dwellings in Westminster, London.
CatBoost was the global-best model, with mean absolute errors of 3.487 SAP and 4.379 EI points on held-out dwellings.
Both strategies projected similar savings; under global-best, roof insulation saved most per dwelling, \pounds203 annually (uncertainty range 176--229), and wall insulation most across the portfolio, \pounds6.09 million annually (uncertainty range 5.69--6.72).
For 5,382 dwellings needing all three upgrades, every order delivered the same final package, yet installing roofs, walls, then windows a year apart accumulated 54\% more projected cost savings over three years than the reverse order.
Emission reductions followed the same pattern.
Using only certificate data, the framework helps local authorities rank measures for dwellings and portfolios and sequence upgrades to bring savings forward.
\end{abstract}

\section{Introduction}

Buildings accounted for 21\% of global greenhouse gas emissions in 2019, mostly through the energy used to operate them \citep{cabeza2022buildings}. In the United Kingdom, the buildings and product uses sector produced 22\% of net territorial greenhouse gas emissions in 2025, and residential buildings produced 66\% of that sector's emissions \citep{desnz2026provisional}. Four out of five homes that will be occupied by 2050 have already been built \citep{ccc2019housing}, which makes improving existing dwellings central to reducing these emissions. Wall insulation, roof insulation and better glazing reduce heat loss, so less energy is needed to keep a home warm. These benefits accrue dwelling by dwelling, while programmes determine which measures receive funding, how widely they are delivered and in what order. Building-level evidence can connect these choices with wider energy, equity and health objectives \citep{chen2025retrofits}. Planning such programmes therefore depends on the benefit each measure is expected to deliver and on an indication of its reliability, since a single value does not show how much higher or lower the saving might be.

That benefit also changes with the scale and timing of a programme. A measure with a large benefit for one dwelling may contribute less across a stock if few dwellings need it. Where several measures are needed, the same final package may deliver different accumulated benefits depending on which improvement comes first. Previous studies have examined how urban context influences retrofit performance through hybrid simulation and data-driven modelling \citep{nutkiewicz2021urbancontext} and incorporated measures and timing into building-stock trajectories \citep{rit2024shape}. A data-driven study of retrofit decisions for the Dublin residential stock drew on a building stock database, noting that developing such databases is time-intensive and that the available data are sparse and inconsistent \citep{ali2020datadriven}.

Energy performance certificates (EPCs) provide standardised dwelling-level data at scale. They were introduced in the European Union to inform the building sector about the energy efficiency of buildings and have become a core source of information on building energy \citep{pasichnyi2019epc}. In England and Wales, a certificate is required when a home is built, sold or let \citep{mhclg2026epc}, and about 1.6 million domestic certificates were lodged in England in the year to March 2026 \citep{mhclg2026epcstatistics}. Each certificate describes the dwelling's walls, roof, windows, floor, heating, hot water and lighting \citep{mhclg2026epcdatadictionary}. It also reports two scores calculated with the Standard Assessment Procedure (SAP) under standardised assumptions: the SAP score reflects modelled energy cost and the environmental impact (EI) score modelled operational carbon dioxide emissions, with higher scores indicating better performance \citep{bre2014sap2012}. A model trained to reconstruct the registered scores from these descriptions can also score an edited dwelling: replacing an uninsulated wall description with an insulated one represents a wall upgrade, and the change in predicted scores gives its projected effect.

Prediction from building records and uncertainty estimation both have established precedents. Neural networks have predicted heating performance from Italian certificates, with fitted ensembles supplying prediction ranges \citep{khayatian2016neural}; symbolic regression has predicted residential energy consumption in Germany \citep{wenninger2022qlattice}. For England and Wales, the Office for National Statistics combined administrative attributes and nearby similar properties to predict missing EPC scores \citep{ons2022predicting}. A gated multimodal study used tabular, textual and geographical information for Westminster score prediction and fabric scenarios \citep{bai2026gated}. Building-energy uncertainty studies have applied conformalised quantile regression to residential-community loads \citep{massidda2023probabilistic} and propagated correlated envelope-parameter uncertainty into dwelling and district heat demand \citep{guo2025uncertainty}. These approaches motivate reporting uncertainty alongside an estimate.

Estimating a retrofit benefit involves a further step: the benefit compares two uncertain states of the same dwelling. Separate prediction intervals do not specify how the errors before and after an edit move together, and therefore do not determine the uncertainty of their difference. Conformal calibration can support marginal coverage of current scores under exchangeability \citep{lei2018distribution}, but it does not supply the unobserved dependence between states. This study makes that dependence explicit through a shared-rank assumption: a score high within its baseline range remains high within its edited range. It also assumes that an idealised fabric upgrade does not worsen either score. Such a physical direction can be built into a predictor \citep{dinatale2022physically}; here it is imposed jointly on the predicted scores of each dwelling's baseline and edited states.

\begin{figure}[!htbp]
\centering
\includegraphics[width=\linewidth]{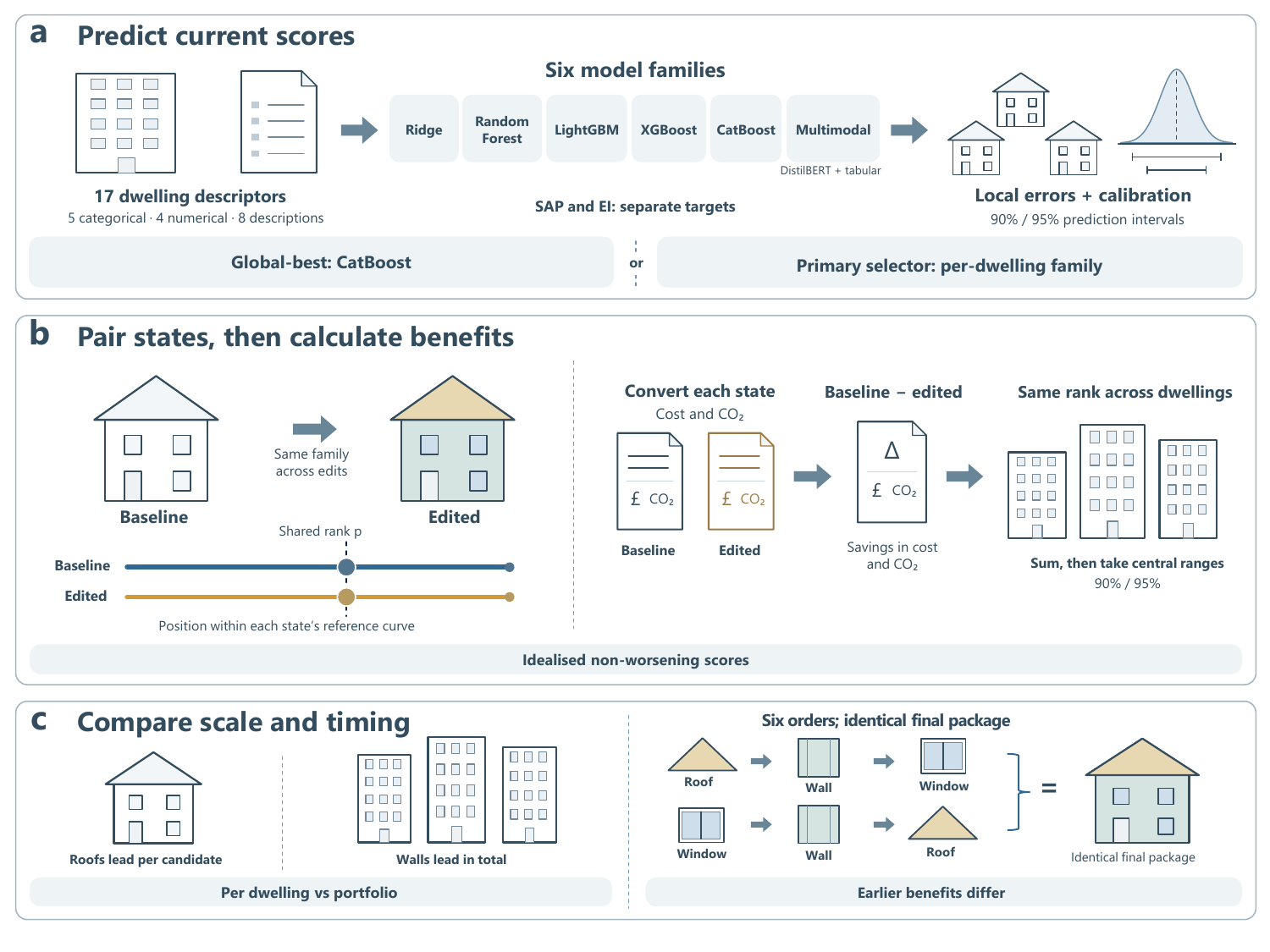}
\caption{From dwelling descriptions to the scale and timing of projected retrofit benefits. \textbf{a}, Six families predict SAP and EI independently from the same descriptors; local out-of-sample errors and independent calibration supply current-score prediction intervals. Global-best and the primary selector provide two choices of predictor. \textbf{b}, The baseline-selected family remains fixed across edits. Shared rank means the same position within each state's reference curve, shown schematically by linked sliders. Joint monotonicity represents idealised non-worsening upgrades. States are converted before subtraction; a further common rank across dwellings defines portfolio summation before central ranges are taken. \textbf{c}, Roofs lead in mean benefit per candidate, whereas walls lead in portfolio totals. Two example orders illustrate different earlier benefits despite an identical completed package; all six orders are evaluated under a common schedule. Building icons and sliders are schematic. Current-score coverage is evaluated against held-out scores; scenario ranges rest on declared assumptions.}
\label{fig:workflow}
\end{figure}

We use 17 certificate descriptors of 103,414 dwellings in Westminster, central London, to examine the projected benefits, reliability and implementation order of wall, roof and window upgrades. The study makes three contributions (Fig.~\ref{fig:workflow}):
\begin{enumerate}[label=(\arabic*),leftmargin=*,itemsep=3pt,topsep=4pt]
\item \textbf{A common empirical reference for score reconstruction with uncertainty.} Six model families are compared on the same building-grouped splits for accuracy, fitting cost and independently calibrated prediction intervals. Paired evaluation tests whether sample-specific selection improves on a development-selected global-best model, while distinguishing predictive accuracy from supported reach.
\item \textbf{An explicit construction for paired retrofit scenarios.} Coherent fabric edits retain each dwelling's baseline-selected model, with joint monotonicity and declared shared-rank assumptions connecting baseline and edited scores. Converting each state before subtraction and summing paired realisations before taking quantiles defines cost and emission ranges at dwelling and portfolio scales.
\item \textbf{A joint assessment of benefit magnitude, applicability and timing.} Common-support comparisons separate projected benefit per candidate from portfolio totals and evaluate all six orders of the same three-measure package. For certificates carrying a single recommended improvement, the rating that the official calculation assigns after that improvement provides a separate reference for mean changes, dwelling-level agreement and scenario ranges.
\end{enumerate}

The remainder of the paper is organised as follows. Section~\ref{sec:prediction-methods} describes the Westminster data, the six model families and the construction and calibration of the prediction intervals, and Section~\ref{sec:scenario-analysis} defines the fabric edits, the pairing of baseline and edited states, their conversion into cost and emission benefits, the staged orders and the comparison with the register's improvement calculations. The results in Section~\ref{sec:results} first evaluate current-score accuracy, interval coverage and model selection, and then compare projected benefits per candidate dwelling and across the portfolio, the accumulated benefits of the six orders of one package, and the agreement between projected and register changes. Section~\ref{sec:discussion} considers what these comparisons imply for retrofit programmes and the limits of the evidence, and Section~\ref{sec:conclusion} concludes.

\FloatBarrier

\section{Methods: score prediction with uncertainty}
\label{sec:prediction-methods}

Six regression families predict the current SAP and EI scores of registered Westminster dwellings from their recorded descriptions, and each family's errors on dwellings with similar descriptions give an interval around each of its predictions (Fig.~\ref{fig:workflow}a). Two strategies then decide which family supplies the prediction for each dwelling.

\subsection{Study area, population and data separation}
\label{sec:cohort}

\begin{figure}[!htbp]
\centering
\includegraphics[width=\linewidth]{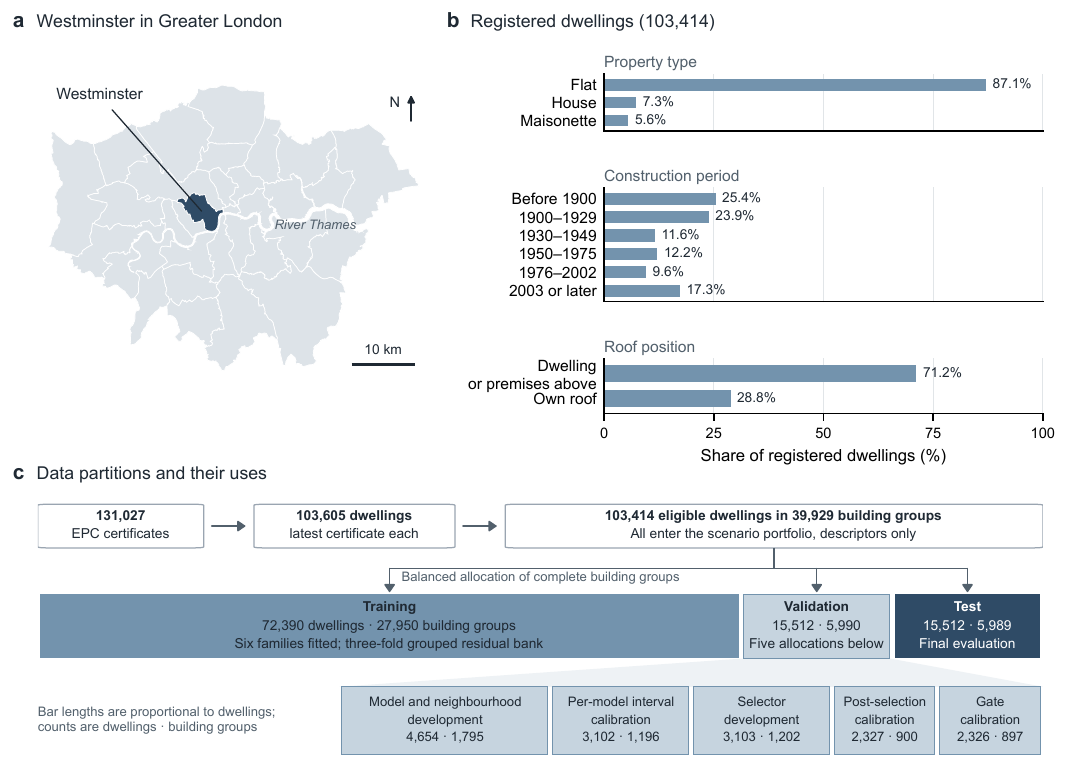}
\caption{Study area, registered dwellings and the use of each data partition. \textbf{a}, Westminster (dark) among the 33 Greater London local authority areas; boundaries clipped to mean high water leave the tidal Thames as a gap. \textbf{b}, Property type, recorded construction period and roof position of the 103,414 registered dwellings. Year-only age records are placed by their year; 39 bungalows or unrecorded property types and 6 unknown or invalid age records are not shown. Roof position separates descriptions of another dwelling or premises above from the dwelling's own roof. \textbf{c}, From certificates to eligible dwellings, building-grouped partitions and the five validation allocations. Bar lengths are proportional to dwellings, and counts give dwellings and building groups. All eligible dwellings, including test records, enter the scenario portfolio through their descriptors only. Boundaries: \citet{ons2025ladboundaries}, Open Government Licence v3.0; contains OS data \textcopyright{} Crown copyright and database right 2025.}
\label{fig:study-area}
\end{figure}

Westminster is a local authority in central London (Fig.~\ref{fig:study-area}a). Among its registered dwellings, 87.1\% are flats and 71.2\% have roof descriptions indicating another dwelling or premises above (Fig.~\ref{fig:study-area}b).

The Westminster preparation retained 103,414 dwellings in 39,929 building groups from 131,027 EPC records (Fig.~\ref{fig:study-area}c). Dwelling identity combined the unique property reference number (UPRN) with unit and address evidence. The latest certificate was selected before cleaning; an invalid latest record removed the dwelling without fallback. Cleaning enforced plausible scores, floor area, room counts and photovoltaic coverage, and excluded negative recorded energy costs \citep{mhclg2019epcglossary}. The Supplementary Information gives every range and the treatment of missing values.

Complete building groups were allocated deterministically to training, validation and test partitions targeting 70\%, 15\% and 15\% of dwellings, so that each test dwelling belongs to a building group unseen during model development (Fig.~\ref{fig:study-area}c). Groups were taken in descending order of size and each was assigned to the partition that best matched the targets for dwelling count, SAP and EI rating-band counts, property type and recorded construction-age band; single-group moves and pairwise swaps then refined the balance. The resulting counts were 72,390, 15,512 and 15,512; the test set contained 5,989 building groups. The same procedure divided the validation set into five disjoint building-group allocations for model and neighbourhood development, per-model calibration, selector development, post-selection calibration and gate calibration for the automatic acceptance of predicted scores (Supplementary Information). Test scores entered only the allocation's band balance and the final evaluation.

\subsection{Common descriptors and six regression families}
\label{sec:point-models}

All six families used the same inputs and predicted the same targets. For dwelling $i$, the descriptor vector $x_i$ contains 17 fields: construction-age band, property type, built form, energy tariff and main fuel; floor area, habitable rooms, heated rooms and photovoltaic coverage; and descriptions of walls, windows, floor, roof, main heating, heating controls, hot water and lighting. The target $y_{i,t}$ is its registered score for $t\in\{\SAP,\EI\}$, and SAP and EI were fitted, selected and calibrated independently throughout.

The families differed in how they represented these fields. Ridge, Random Forest \citep{breiman2001forests}, LightGBM \citep{ke2017lightgbm} and XGBoost \citep{chen2016xgboost} one-hot encoded all 13 string fields, including complete description strings. CatBoost treated these strings as native categorical features \citep{prokhorenkova2018catboost}. Multimodal combined a fully fine-tuned DistilBERT text representation with categorical embeddings and numerical features through a learned fusion gate \citep{sanh2019distilbert}. Validation MAE selected each configuration; predictor equations, architecture, grids and settings are in the Supplementary Information. Reported predictions were $f_{m,t}(x)=\clip\{f^{\mathrm{raw}}_{m,t}(x)\}$. A raw prediction outside 1--100 made that candidate ineligible for selection.

\subsection{Local prediction intervals and independent calibration}
\label{sec:local-interval}

For a new dwelling, the interval construction asks how far a given model has erred on similar descriptions. Similarity was model-specific: Ridge used train-standardised contributions at the 17-field level, tree families used terminal-leaf co-occurrence, and Multimodal used its normalised fusion embedding. A three-fold building-grouped residual bank supplied signed errors $e_{j,m,t}=y_{j,t}-f^{[-\mathrm{fold}(j)]}_{m,t}(x_j)$ for the 72,390 training dwellings. Each residual came from a fit excluding the dwelling and its building group.

The $K$ nearest reference descriptions, denoted $\mathcal N_K(x)$, supplied distance-weighted residual quantiles. Let $d_j(x)$ be the distance to neighbour $j$ and $d_K(x)$ the largest retrieved distance. All selected configurations used adaptive Gaussian weights, with effective neighbour count
\begin{equation}
 \begin{aligned}
 w_j(x)&=\exp\!\left[-\frac{d_j(x)^2}{\tau(x)^2}\right],
 &\tau(x)&=\max\{d_K(x),10^{-12}\},\\
 N_{\mathrm{eff}}(x)&=\frac{\left[\sum_{j\in\mathcal N_K(x)}w_j(x)\right]^2}
 {\sum_{j\in\mathcal N_K(x)}w_j(x)^2}.
 \end{aligned}
 \label{eq:local-weights}
\end{equation}
Model and target indices are suppressed in these neighbourhood quantities. Effective count measures how many reference residuals contribute after weighting. At nominal level $1-\alpha$, the base interval was
\begin{equation}
 \widetilde C_{m,t}^{1-\alpha}(x)=
 [f_{m,t}(x)+Q_{\alpha/2}^{w}(e\mid x),\;
  f_{m,t}(x)+Q_{1-\alpha/2}^{w}(e\mid x)].
 \label{eq:base-interval}
\end{equation}
For 90\%, these are the weighted 5th and 95th residual percentiles; 95\% uses the 2.5th and 97.5th percentiles. Signed errors permit asymmetric bounds. Neighbourhood size and weighting were chosen on development data. Effective neighbour count, distance and residual spread determined local support. Failed support invoked global residual quantiles and excluded the candidate from model selection. All definitions and fixed thresholds are in the Supplementary Information.

Local quantiles alone do not establish nominal coverage. Each model, target and level therefore received an independent conformal correction. On its calibration allocation, nonconformity was
\begin{equation}
 \xi_i^\alpha=\max\{\widetilde L_i^\alpha-y_i,\;y_i-\widetilde U_i^\alpha,\;0\}.
 \label{eq:nonconformity}
\end{equation}
The correction $q^\alpha$ was the order statistic at rank $\lceil(n_{\mathrm{cal}}+1)(1-\alpha)\rceil$, applied as
\begin{equation}
 C^{1-\alpha}(x)=[\clip\{\widetilde L^\alpha(x)-q^\alpha\},\;
                    \clip\{\widetilde U^\alpha(x)+q^\alpha\}].
 \label{eq:calibrated}
\end{equation}
Corrections were fitted separately at 90\% and 95\%; intervals are reported at 90\%, with 95\% alongside. Under exchangeability and an independently fixed base procedure, this construction supplies marginal coverage, not a guarantee conditional on every dwelling \citep{lei2018distribution,barber2021limits}.

\subsection{Global-best reference and sample-specific selection}
\label{sec:selection}

Global-best chooses one family for each target by its MAE on the selector-development allocation and uses it for every dwelling. CatBoost was selected for both targets. The primary selector instead chooses an eligible family separately for each dwelling and target. Eligibility requires finite outputs, ordered bounded intervals, an in-range raw prediction and adequate local support. A strategy routes a dwelling when it returns a supported prediction; global-best leaves a dwelling unrouted if CatBoost is ineligible, even when another family is eligible.

The primary selector combines weighted local absolute residuals with MAE on the selector-development allocation, denoted $\mathrm{MAE}^{\mathrm{dev}}_{m,t}$. Its error estimate and local weight are
\begin{align}
 \widehat R_{m,t}(x)
 &=[1-\rho(x)]\mathrm{MAE}^{\mathrm{dev}}_{m,t}
 +\rho(x)\frac{\sum_{j\in\mathcal N_K(x)}w_j(x)|e_{j,m,t}|}
 {\sum_{j\in\mathcal N_K(x)}w_j(x)},\label{eq:shrunk-risk}\\
 \rho(x)&=\frac{N_{\mathrm{eff}}(x)}{N_{\mathrm{eff}}(x)+\kappa_t}
 \max\!\left\{0,1-\frac{d_K(x)}{d_{K,\max}}\right\}.
 \label{eq:local-reliability}
\end{align}
Here $d_{K,\max}$ is the model- and target-specific development distance limit used for support qualification, and $\kappa_t$ controls shrinkage. The model and target indices of $\rho$ and the neighbourhood quantities are suppressed. Fewer effective neighbours or a query nearer the distance limit reduce the local weight, moving the estimate towards development MAE. For eligible set $\mathcal E_t(x)$, the selector chooses
\begin{equation}
 m_t^*(x)=\arg\min_{m\in\mathcal E_t(x)}
 \left\{\widehat R_{m,t}(x)+\lambda_t
 \frac{U_{m,t}^{0.95}(x)-L_{m,t}^{0.95}(x)}2\right\}.
 \label{eq:selector}
\end{equation}
Five building-grouped development folds selected $\kappa_t$ and $\lambda_t$ using a loss that penalised absolute errors above five points. The Supplementary Information gives the selected values and the tie-breaking rule, and defines three pre-specified alternatives evaluated alongside the primary selector: a width-free selector, a learned router and a soft ensemble.

Selection changes the distribution of reported intervals. Every target, strategy and level, including global-best, therefore received a further independent correction on its routed post-selection allocation using Equation~\eqref{eq:nonconformity}.

\subsection{Evaluation of accuracy, coverage and paired selection}
\label{sec:evaluation}

Point accuracy was evaluated by MAE, RMSE and $R^2$, with MAE determining the principal ranking. Computational cost was the recorded time of each family's configuration search and fitting. Interval evaluation reported coverage, width and the proper interval score, which penalises both width and uncovered outcomes \citep{gneiting2007proper}. Each family's intervals were evaluated on all test dwellings, including global fallback intervals where support failed, and each strategy's intervals on its own routed dwellings. Property type, construction age and main fuel defined pre-specified subgroups; a reported entry required at least 100 selected dwellings and 50 building groups.

Selection accuracy was compared on each strategy's own routed dwellings, using the same dwellings for its predictions and CatBoost's. For routed set $\mathcal Q_{s,t}$, strategy prediction $\widehat y_{s,t}$ and global-best family $m_t^{\mathrm{GB}}$, paired MAE gain was
\begin{equation}
 \Delta^{\mathrm{MAE}}_{s,t}=\frac1{|\mathcal Q_{s,t}|}
 \sum_{i\in\mathcal Q_{s,t}}
 \left[|y_{i,t}-f_{m_t^{\mathrm{GB}},t}(x_i)|
       -|y_{i,t}-\widehat y_{s,t}(x_i)|\right].
 \label{eq:paired-gain}
\end{equation}
A positive gain favours the strategy. CatBoost point predictions entered this paired comparison even when they failed its own support rule. A full-pool outcome-informed oracle, which takes for each dwelling the prediction with the smallest observed error among all six families, gives the lowest error attainable by selecting one of these predictions. MAE and paired gain received 95\% confidence intervals from 1,000 building-group bootstrap replicates, which resampled complete groups and retained dwelling weighting.

SAP 2012 rating bands provided an additional reporting unit \citep{bre2014sap2012}. Predictions and interval endpoints were rounded to integer ratings before banding. Band agreement, interval spans, complete metric definitions and numerical results are in the Supplementary Information.

\section{Model-based retrofit scenario analysis}
\label{sec:scenario-analysis}

A retrofit benefit compares the same dwelling before and after a change to its fabric (Fig.~\ref{fig:workflow}b). The family selected for each dwelling scores both its recorded description and edited descriptions representing wall, window and roof upgrades, and the two states are paired under declared physical and shared-rank assumptions. The paired scores are converted into annual cost and emission benefits, compared per dwelling, across the portfolio and between implementation orders, and checked against the register's improvement calculation.

\subsection{Coherent edits and fixed predictors}
\label{sec:scenario-design}

The scenario portfolio comprised all 103,414 registered Westminster dwellings, across the training, validation and test partitions. Fitted models supply the edited scores because the official Reduced Data SAP (RdSAP) calculation cannot be repeated from the open register, which omits several of its required inputs, including detailed geometry, window orientation and appliance information \citep{bre2019rdsap,mhclg2026epcdatadictionary}. The Supplementary Information lists the field-level gap.

A coarse screen of the wall, window and roof descriptions identified the candidates for each upgrade. Wall and roof edits changed explicitly absent, partial or limited insulation to an insulated state, and roof edits applied only to dwellings without another dwelling or premises above. Window edits upgraded single or secondary glazing to full double glazing and completed partial double or higher glazing. Each transformation $T_a$ changed only the relevant field, preserving construction type and textual qualifiers. For example, \emph{Solid brick, as built, no insulation (assumed)} became \emph{Solid brick, as built, insulated (assumed)}. Every edited description was checked for local support and against the training vocabulary; a description absent from that vocabulary is an unseen category for the five families that encode complete descriptions as categories.

Both strategies were carried through all scenarios. Global-best used CatBoost; the primary selector chose a family on the baseline description and kept it fixed across every applicable edit and combination. For strategy $s$, target $t$ and dwelling $i$, denote this family by $m_{s,t,i}$, which may differ between SAP and EI, and its state predictions by
\begin{equation}
 \widehat y_{i,t}^{S,s}=f_{m_{s,t,i},t}\{T_S(x_i)\},\qquad
 S\subseteq\mathcal A_i,
 \label{eq:scenario-contrast}
\end{equation}
where $\mathcal A_i$ contains the applicable measures and $T_\varnothing(x_i)=x_i$. Fixed model identity prevents a descriptor response from being mixed with a model-switch response. Edited states received intervals from the same representations, residual banks, support thresholds and calibration corrections as the baseline.

\subsection{Pairing uncertain states under explicit assumptions}
\label{sec:paired-scenario}

The physical scenario assumes correctly installed upgrades under fixed floor area, heating system, fuel, service demand, reference prices and emission factors. Scores may stay unchanged but may not worsen as measures are added. For glazing, this additionally assumes that altered solar gains or ventilation do not offset the insulation benefit \citep{bre2014sap2012}. To impose this weak order coherently across combinations, edited scores were projected jointly while keeping the baseline fixed:
\begin{equation}
 \bar{\boldsymbol z}=\arg\min_{\boldsymbol z}
 \sum_{\varnothing\ne S\subseteq\mathcal A_i}(z_S-\widehat y^S)^2,
 \quad
 \begin{gathered}
 z_\varnothing=\widehat y^\varnothing,\quad 1\leq z_S\leq100,\\
 z_S\leq z_T\quad\text{whenever }S\subseteq T.
 \end{gathered}
 \label{eq:monotone-projection}
\end{equation}
Indices $i,t,s$ are suppressed. Dykstra's iterative projections solved this equal-weight least-squares problem \citep{dykstra1983restricted}. Because any applicable subset can influence the joint projection, comparisons were restricted to common support: dwellings supported at the baseline and at every applicable subset under both targets and both strategies.

The dependence assumption can be pictured as linked sliders. Each state has a reference curve through its lower 95\%, lower 90\%, point, upper 90\% and upper 95\% predictions. A shared slider position $p$ selects the same rank within every state's curve: a score high in its baseline range is paired with a score high in its edited range. This preserves shared variation when forming a difference.

The curves place these predictions at ranks $0.025$, $0.05$, $0.5$, $0.95$ and $0.975$, with linear interpolation and bounded tails. At each of 401 equally weighted ranks, the full state vector is projected by Equation~\eqref{eq:monotone-projection}, anchoring the baseline to its value at that rank; write the result as $z_S(p)$. An unchanged description then gives zero benefit at every rank. The Supplementary Information details the projection and the reference curves.

\subsection{From paired scores to dwelling and portfolio benefits}
\label{sec:conversion}

SAP 2012 version 9.92 supplies a common conversion convention \citep{bre2014sap2012}. For floor area $A_i$, let $C(y,A_i)$ be annual rating-equivalent cost in pounds and $E(y,A_i)$ annual operational emissions in kilograms of CO$_2$, obtained by inverting the SAP and EI relationships respectively. The Supplementary Information gives the piecewise mappings.

For either decreasing mapping $g\in\{C,E\}$ and its corresponding target, projected point benefits and paired realisations are
\begin{align}
 \widehat B_{g,i}^{S,s}
 &=g(\bar z_i^{\varnothing,s},A_i)-g(\bar z_i^{S,s},A_i),\label{eq:benefits}\\
 B_{g,i}^{S,s}(p)
 &=g\{z_i^{\varnothing,s}(p),A_i\}-g\{z_i^{S,s}(p),A_i\}.\label{eq:paired-benefit}
\end{align}
Both states are converted before subtraction because the mapping can be nonlinear. Nonnegative benefits follow from the imposed score order.

Portfolio aggregation adds a second dependence assumption: all dwellings share the same rank $p$, separately for each target and strategy. Their sliders therefore move together across the portfolio as well as within each dwelling. For a fixed candidate set $\mathcal P_a$, summation precedes the calculation of central ranges:
\begin{equation}
 B_{g,\mathcal P_a}^{a,s}(p)=\sum_{i\in\mathcal P_a}B_{g,i}^{\{a\},s}(p),
 \qquad
 \mathcal R_{1-\alpha}=[Q_{\alpha/2}\{B(p)\},Q_{1-\alpha/2}\{B(p)\}].
 \label{eq:portfolio}
\end{equation}
Quantiles use the fixed rank grid, and central ranges are reported at 90\%, with 95\% alongside. Dividing a total and its range by the number of candidates gives the per-candidate mean and its range. Table~\ref{tab:evidence-status} distinguishes the interpretation of these ranges from that of prediction intervals and bootstrap confidence intervals.

\begin{table}[!htbp]
\centering\small
\caption{Interpretation of the reported uncertainty quantities.}
\label{tab:evidence-status}
\begin{tabular}{L{3.5cm}L{10.7cm}}
\toprule
Quantity & Interpretation\\
\midrule
Prediction interval & A current registered score; marginal coverage assessed on held-out dwellings, without a guarantee for every dwelling or subgroup.\\
Central scenario range & Variation under the declared physical and shared-rank assumptions; no validated coverage of realised retrofit effects or portfolio totals.\\
Bootstrap confidence interval & Sampling uncertainty in a summary statistic, from building-group resampling conditional on fitted models.\\
\bottomrule
\end{tabular}
\end{table}

\subsection{The same package delivered in different orders}
\label{sec:sequences}

Two measures yield two orders and three measures yield six. For order $\pi=(a_1,\ldots,a_k)$, let $S_j=\{a_1,\ldots,a_j\}$. The edits affect different fields and commute, so identical final descriptions give identical final predictions. Every order therefore ends with the same annual package benefit. Intermediate states can differ, changing when savings arrive.

If state $S_j$ persists for $\Delta t_j$ years, accumulated benefit is
\begin{equation}
 H_{g,i}^{\pi,s}=\sum_{j=1}^{k}\Delta t_j
 [g(\bar z_i^{\varnothing,s},A_i)-g(\bar z_i^{S_j,s},A_i)].
 \label{eq:staged-benefit}
\end{equation}
The illustrative schedule assigns one year to each state. All orders of a package are compared on the same dwellings: those on common support that need exactly that package. Pairwise order contrasts are formed within each shared-rank realisation before aggregation and quantiles. The Supplementary Information decomposes their point differences into timing and non-additivity of intermediate responses.

\subsection{Checking projected changes against the register's engine}
\label{sec:engine-comparison}

The register's potential ratings are the official engine's ratings after its recommended improvements. RdSAP Appendix~T applies these cumulatively by editing the assessor's inputs and recalculating \citep{bre2019rdsap,mhclg2026epcdatadictionary}. A certificate carrying exactly one corresponding recommendation therefore supplies an engine-computed single-measure score change: potential minus current rating. The comparison used coarse-screen candidates on common support whose certificates carried exactly one distinct recommendation identifier in the corresponding wall, window or roof family. The Supplementary Information lists the included identifiers.

For each measure, target and strategy, let $\mathcal V$ contain the matched certificates and $n=|\mathcal V|$. Suppressing these comparison indices, the register change, constrained projected change and paired difference were
\begin{equation}
 \begin{aligned}
 \Delta_i^{\mathrm{reg}}&=y_i^{\mathrm{potential}}-y_i^{\mathrm{current}},
 &\Delta_i^{\mathrm{proj}}&=\bar z_i^{\{a\}}-\bar z_i^{\varnothing},\\
 d_i&=\Delta_i^{\mathrm{proj}}-\Delta_i^{\mathrm{reg}}.
 \end{aligned}
 \label{eq:engine-difference}
\end{equation}
Table~\ref{tab:engine-validation} reports the two mean changes and three distinct summaries of these paired differences:
\begin{equation}
 \begin{aligned}
 \overline d&=\frac1n\sum_{i\in\mathcal V}d_i,
 &\mathrm{Mean}\,|\mathrm{Diff.}|&=\frac1n\sum_{i\in\mathcal V}|d_i|,\\
 p_5&=\frac1n\sum_{i\in\mathcal V}\ind\{|d_i|\leq5\}.
 \end{aligned}
 \label{eq:engine-metrics}
\end{equation}
The first measures signed mean disagreement and the second mean absolute disagreement across dwellings; the third is the share within five points, shown as a percentage under $|\mathrm{Diff.}|\leq5$. Mean-difference 95\% confidence intervals used 1,000 building-group bootstrap replicates. Agreement across dwellings is further described by Spearman correlation and by the share of register changes inside each dwelling's central 90\% and 95\% ranges, both given in the Supplementary Information.

\FloatBarrier
\section{Results}
\label{sec:results}

On held-out test dwellings, CatBoost reconstructed registered scores most accurately, choosing a family for each dwelling extended support without improving accuracy, and the recalibrated intervals of both strategies reached nominal marginal coverage. Under both strategies, the projected benefits ranked roof insulation first per candidate dwelling and wall insulation first across the portfolio, and reordering a three-measure package changed the benefit that accrued before it was complete (Fig.~\ref{fig:workflow}c). The projections agreed with the register's improvement calculation more closely in their mean changes than dwelling by dwelling.

\subsection{Current-score evaluation established the prediction reference}
\label{sec:dwelling-results}

Every projected benefit is calculated from predicted scores, so the scenario results rest on how accurately the six families and the two strategies reconstruct the current scores of held-out test dwellings, and on how well their intervals cover those scores.

\subsubsection{Global-best accuracy and fitting resources}

\begin{figure}[!htbp]
\centering
\includegraphics[width=\linewidth]{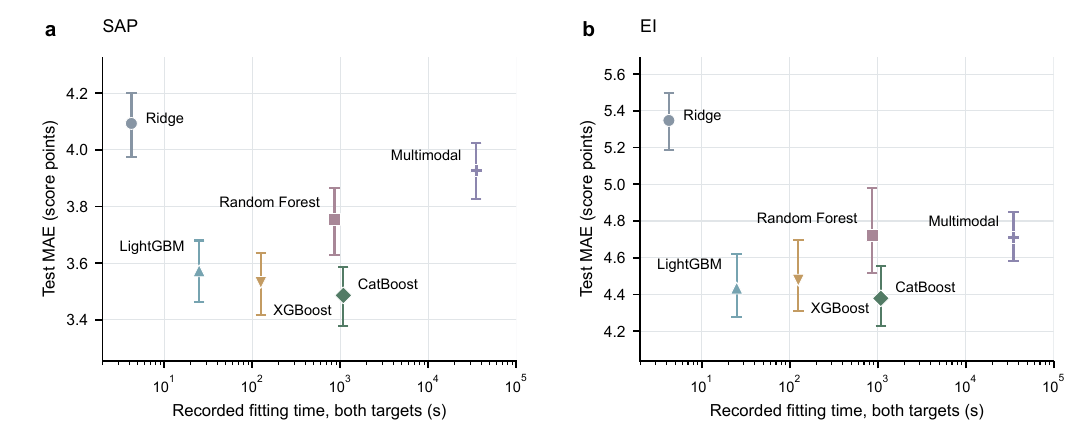}
\caption{Westminster point accuracy, sampling uncertainty and recorded fitting time. \textbf{a}, SAP and \textbf{b}, EI test MAE for all six families on the same 15,512 dwellings in 5,989 building groups. Points show estimated MAE; error bars show 95\% confidence intervals for each family's MAE from 1,000 building-group bootstrap replicates. The logarithmic horizontal axis shows fitting time summed over both targets, including configuration search.}
\label{fig:accuracy-cost}
\end{figure}

CatBoost, selected as global-best during development, also had the lowest observed test MAE for both targets, 3.487 SAP and 4.379 EI points, with 95\% building-group bootstrap confidence intervals of 3.379--3.586 and 4.228--4.554 (Table~\ref{tab:local-point}; Fig.~\ref{fig:accuracy-cost}). XGBoost ranked second for SAP at 3.532 points and LightGBM second for EI at 4.435 points; Ridge had the largest errors for both targets.

Small differences in error accompanied large differences in recorded fitting time. CatBoost required 1,085\,s across both targets, 43.4 times LightGBM's 25\,s, for MAEs lower by 0.087 SAP and 0.056 EI points, and Multimodal required 35,049\,s without attaining the lowest MAE for either target.

\begin{table}[!htbp]
\centering\small
\caption{Point-prediction accuracy on the same 15,512 Westminster test dwellings in 5,989 building groups. Errors are in score points.}
\label{tab:local-point}
\begin{tabular}{lrrrrrr}
  \toprule
  Model & SAP MAE & RMSE & $R^2$ & EI MAE & RMSE & $R^2$ \\
  \midrule
  Ridge & 4.094 & 5.725 & 0.721 & 5.348 & 7.390 & 0.737 \\
  Random Forest & 3.754 & 5.496 & 0.743 & 4.720 & 7.168 & 0.753 \\
  LightGBM & 3.574 & 5.103 & 0.778 & 4.435 & 6.373 & 0.804 \\
  XGBoost & 3.532 & 5.056 & 0.782 & 4.477 & 6.532 & 0.795 \\
  CatBoost & 3.487 & 4.960 & 0.790 & 4.379 & 6.249 & 0.812 \\
  Multimodal & 3.926 & 5.438 & 0.748 & 4.713 & 6.574 & 0.792 \\
  \bottomrule
\end{tabular}

\end{table}

\subsubsection{Local interval coverage and precision}

\begin{figure}[!htbp]
\centering
\includegraphics[width=\linewidth]{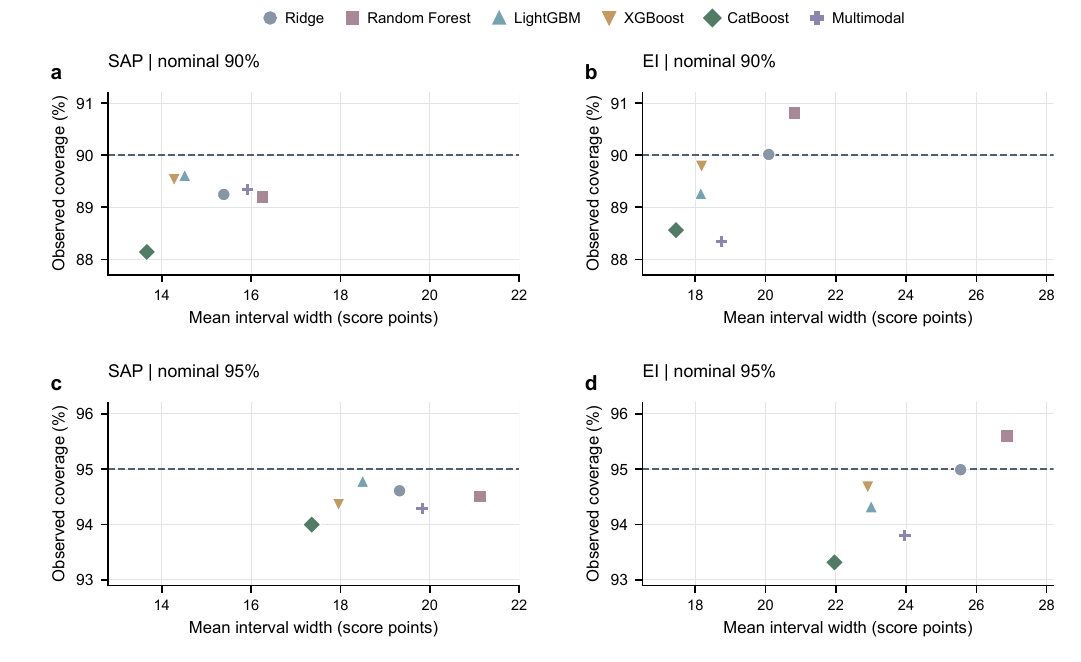}
\caption{Empirical current-score coverage in relation to interval width. \textbf{a,b}, Independently calibrated nominal 90\% intervals for SAP and EI. \textbf{c,d}, Independently calibrated nominal 95\% intervals. Each symbol represents one family evaluated on all 15,512 Westminster test dwellings, including global fallback intervals for unsupported descriptions. Dashed lines mark nominal coverage; the horizontal axis is mean width in score points.}
\label{fig:interval-reliability}
\end{figure}

Each family's own intervals fell slightly short of nominal coverage for SAP, and the narrowest did so for both targets (Fig.~\ref{fig:interval-reliability}). CatBoost's 90\% intervals had the smallest mean widths, 13.67 SAP and 17.45 EI points, and covered 88.1\% and 88.6\% of registered test scores. Across the six families, 90\% coverage ranged from 88.1\% to 89.6\% for SAP, so all six fell below nominal, and from 88.3\% to 90.8\% for EI, where four of the six fell below. Random Forest's higher EI coverage came with a larger mean width, 20.82 points, and a worse mean 90\% interval score, 30.18 against CatBoost's 26.50, so coverage alone did not rank the families. For each dwelling, the 90\% interval score is the interval's width plus twenty times any distance by which the registered score falls outside it, so lower mean scores indicate a better combined width--miscoverage trade-off. The 95\% intervals showed the same pattern, with all six SAP families at 94.0--94.8\% (Fig.~\ref{fig:interval-reliability}c,d).

The unweighted mean of the six model-specific support-failure rates was 5.1\% for SAP and 5.3\% for EI; another family could still supply a supported prediction when one candidate failed its support checks.

\subsubsection{Sample-specific selection and recalibrated intervals}

\begin{figure}[!htbp]
\centering
\includegraphics[width=\linewidth]{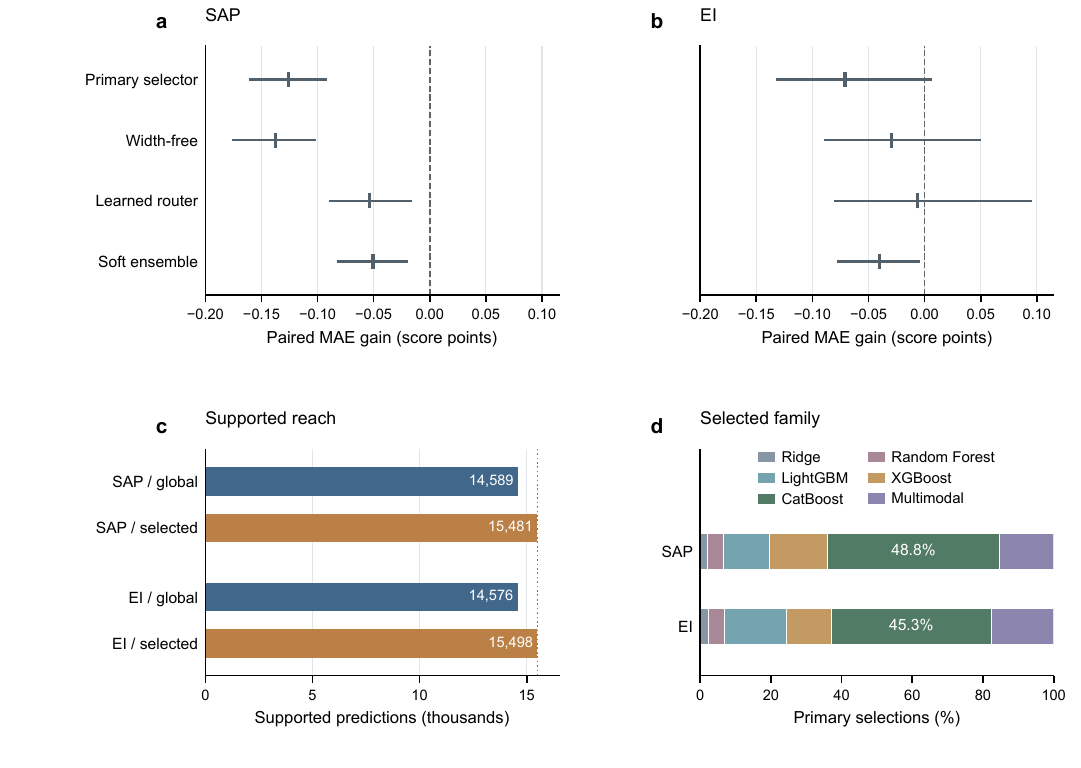}
\caption{Paired accuracy and supported reach of the pre-specified selection strategies. \textbf{a,b}, MAE gains relative to the global-best family, CatBoost, evaluated on each strategy's own matched routed dwellings. Vertical ticks mark estimated gains; horizontal lines show 95\% confidence intervals from 1,000 building-group bootstrap replicates. Positive values favour the strategy. \textbf{c}, Supported prediction counts for global-best and the primary selector; the dotted line marks all 15,512 test dwellings. \textbf{d}, Family composition among 15,481 SAP and 15,498 EI primary selections, with CatBoost shares labelled. The family colour key applies to \textbf{d}.}
\label{fig:selection-tradeoff}
\end{figure}

Choosing a family for each dwelling extended support but did not improve accuracy. The primary selector routed 15,481 SAP and 15,498 EI dwellings, against 14,589 and 14,576 under global-best, which applies CatBoost throughout (Fig.~\ref{fig:selection-tradeoff}c). On exactly the same routed dwellings, its SAP MAE was 3.604 points against CatBoost's 3.478, a paired gain of $-0.126$ points (95\% bootstrap confidence interval $-0.160$ to $-0.092$); its EI MAE was 4.448 against 4.377, a gain of $-0.071$ ($-0.132$ to 0.005).

The three pre-specified alternatives also had negative point gains, every SAP confidence interval lay below zero, and for EI only the soft ensemble's interval lay entirely below zero (Fig.~\ref{fig:selection-tradeoff}a,b). CatBoost supplied 48.8\% of SAP and 45.3\% of EI primary selections. A full-pool outcome-informed oracle attained MAEs of 2.135 SAP and 2.613 EI points on the primary routed dwellings, so other families were sometimes more accurate than CatBoost, but none of the deployable rules recovered this complementarity. CatBoost therefore remains the accuracy reference, and the primary selector is carried into the scenario analysis as a test of whether its conclusions depend on model choice.

\begin{table}[!htbp]
\centering\small
\caption{Independently recalibrated strategy intervals on each strategy's own supported Westminster test dwellings. Inside gives coverage counts and percentages; $W$ is mean width in score points.}
\label{tab:strategy-intervals}
\begin{tabular}{llrrrrr}
  \toprule
  Target & Strategy & $n$ & Inside 90 (\%) & Inside 95 (\%) & $W_{90}$ & $W_{95}$ \\
  \midrule
  SAP & Global-best & 14,589 & 13,540 (92.8) & 14,087 (96.6) & 15.25 & 19.16 \\
  SAP & Primary selector & 15,481 & 14,245 (92.0) & 14,796 (95.6) & 16.02 & 19.26 \\
  \midrule
  EI & Global-best & 14,576 & 13,381 (91.8) & 13,963 (95.8) & 18.90 & 24.09 \\
  EI & Primary selector & 15,498 & 14,110 (91.0) & 14,810 (95.6) & 18.92 & 23.77 \\
  \bottomrule
\end{tabular}

\end{table}

After independent post-selection calibration, both strategies' 90\% intervals reached nominal marginal coverage on their own routed test dwellings (Table~\ref{tab:strategy-intervals}). The primary selector's 90\% intervals covered 92.0\% of SAP and 91.0\% of EI test scores at mean widths of 16.02 and 18.92 points; its 95\% intervals covered 95.6\% of both at 19.26 and 23.77 points. Converting each endpoint at its own floor area through the mappings of Section~\ref{sec:conversion} gives mean 90\% spans of \pounds362.34 per year and 2,033.02\,kg CO$_2$ per year, with medians of \pounds283.06 and 1,533.78\,kg.

Coverage varied across subgroups: of 40 supported observable-subgroup entries, 5 (12.5\%) had selected 90\% coverage below 90\%, with a minimum of 87.7\%; at the 95\% level, 20.0\% fell below nominal, with a minimum of 92.5\%.

\subsection{Roofs led per candidate dwelling and walls across the portfolio}
\label{sec:benefit-results}

\begin{figure}[!htbp]
\centering
\includegraphics[width=\linewidth]{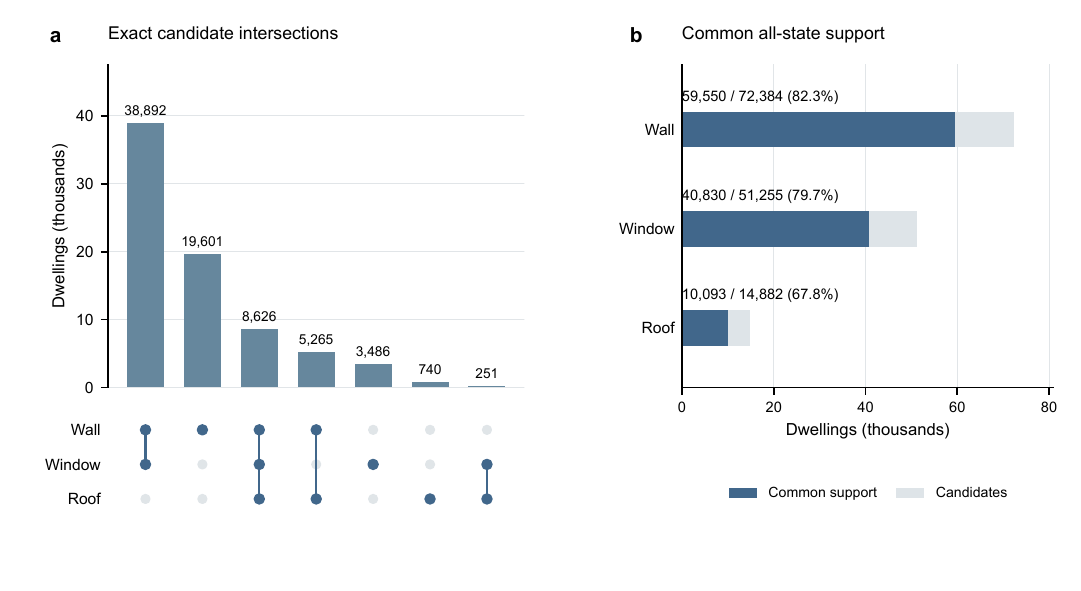}
\caption{Overlapping retrofit candidates and supported comparison populations in the 103,414-dwelling Westminster portfolio. \textbf{a}, Exact, mutually exclusive combinations among the 76,861 dwellings with at least one candidate measure. Filled matrix entries identify each combination; 26,553 further dwellings have no candidate measure. \textbf{b}, Common-support counts as a fraction of each measure's candidate population. Support intersects every applicable edited subset under both targets and strategies, yielding identical comparison populations within each measure. Measures overlap, so their totals are not additive.}
\label{fig:scenario-applicability}
\end{figure}

The first planning comparison separates the size of a dwelling's response from the number of dwellings that can contribute. Screening the 103,414-dwelling portfolio identified 72,384 wall, 51,255 window and 14,882 roof candidates (Fig.~\ref{fig:scenario-applicability}a). Roof exposure limited reach: 73,616 records, 71\% of the portfolio, indicated another dwelling or premises above. At least one measure applied to 76,861 dwellings, 53,034 required two or more and 8,626 required all three.

On common support, the comparison populations were 59,550 wall, 40,830 window and 10,093 roof candidates (Fig.~\ref{fig:scenario-applicability}b). Under its own support rule alone, the primary selector supported more wall candidates than global-best: 67,250 against 62,710 for SAP and 67,830 against 61,447 for EI (Supplementary Information). The common-support comparison therefore separates the projected benefit from differences in which dwellings a strategy can serve.

\begin{figure}[!htbp]
\centering
\includegraphics[width=\linewidth]{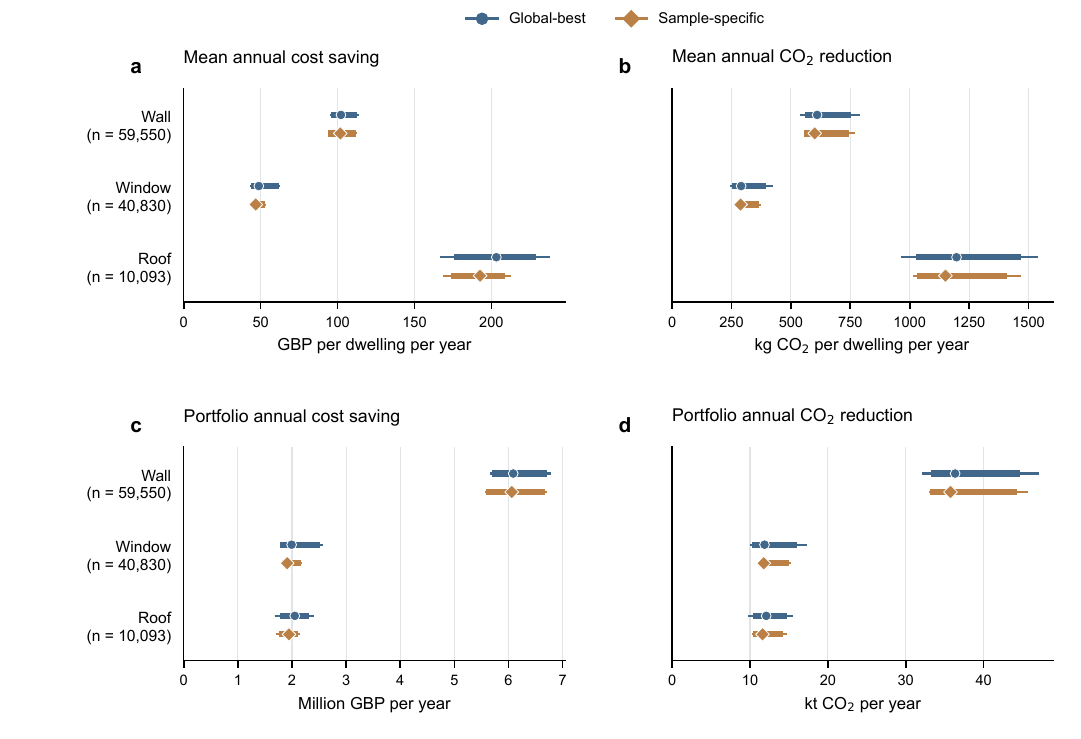}
\caption{Mean candidate benefits and portfolio totals answer different planning questions. \textbf{a,b}, Mean annual standardised energy-cost savings and operational CO$_2$ reductions per candidate dwelling. \textbf{c,d}, Portfolio totals. Symbols show projected point benefits; thick and thin lines show central 90\% and 95\% scenario ranges. Total ranges follow summation within each shared rank; mean ranges divide totals by the fixed population size and do not describe individual dwellings. Comparison populations are identical across strategies and targets within each measure, but differ across measures.}
\label{fig:scenario-benefits}
\end{figure}

Roof insulation gave the largest projected mean benefit per candidate, whereas wall insulation gave the largest portfolio total (Table~\ref{tab:scenario-benefits}; Fig.~\ref{fig:scenario-benefits}). Under global-best, mean standardised annual cost savings were about \pounds203 for roofs, \pounds102 for walls and \pounds49 for windows. Wall insulation was projected to save \pounds6.09 million per year [5.69, 6.72] and reduce operational emissions by 36.34\,kt CO$_2$ per year [33.21, 44.72]. Brackets denote central 90\% scenario ranges. Roof and window cost totals were smaller and similar, \pounds2.05 million [1.78, 2.31] and \pounds1.99 million [1.78, 2.52]: the larger roof response applied to far fewer candidates.

\begin{table}[!htbp]
\centering\footnotesize
\setlength{\tabcolsep}{3pt}
\caption{Single-measure projections on common support. Score changes are property means. Cost saving is in million GBP per year and emission reduction in kt CO$_2$ per year. Brackets contain central 90\% scenario ranges after paired conversion and summation within each shared rank; the Supplementary Information gives both central levels.}
\label{tab:scenario-benefits}
\begin{tabular}{@{}llrrrrr@{}}
\toprule
Measure & Strategy & $n$ & $\Delta$SAP & $\Delta$EI & Cost saving & CO$_2$ reduction \\
\midrule
Wall & Global-best & 59,550 & 4.97 & 6.61 & \shortstack{6.09\\[-1pt]{\scriptsize [5.69, 6.72]}} & \shortstack{36.34\\[-1pt]{\scriptsize [33.21, 44.72]}} \\
Wall & Primary selector & 59,550 & 4.99 & 6.53 & \shortstack{6.06\\[-1pt]{\scriptsize [5.58, 6.67]}} & \shortstack{35.74\\[-1pt]{\scriptsize [33.11, 44.29]}} \\
Window & Global-best & 40,830 & 2.41 & 3.19 & \shortstack{1.99\\[-1pt]{\scriptsize [1.78, 2.52]}} & \shortstack{11.86\\[-1pt]{\scriptsize [10.22, 16.06]}} \\
Window & Primary selector & 40,830 & 2.34 & 3.15 & \shortstack{1.91\\[-1pt]{\scriptsize [1.82, 2.16]}} & \shortstack{11.77\\[-1pt]{\scriptsize [11.19, 14.95]}} \\
Roof & Global-best & 10,093 & 9.68 & 12.50 & \shortstack{2.05\\[-1pt]{\scriptsize [1.78, 2.31]}} & \shortstack{12.08\\[-1pt]{\scriptsize [10.34, 14.81]}} \\
Roof & Primary selector & 10,093 & 9.12 & 11.91 & \shortstack{1.94\\[-1pt]{\scriptsize [1.75, 2.11]}} & \shortstack{11.61\\[-1pt]{\scriptsize [10.39, 14.23]}} \\
\bottomrule
\end{tabular}

\end{table}

The primary selector gave the same two-scale finding, with a wall cost total of \pounds6.06 million [5.58, 6.67]. Within the positive portfolio totals, 15.63\% of global-best window candidates had a zero lower bound in their individual central 90\% ranges. Joint projection changed 3.30--7.83\% of edited point scores, depending on measure, target and strategy, with mean absolute adjustments of 0.008--0.100 points but an individual maximum of 38.57 EI points (Supplementary Information).

\subsection{The same completed package delivered different earlier benefits}
\label{sec:sequence-results}

\begin{figure}[!htbp]
\centering
\includegraphics[width=\linewidth]{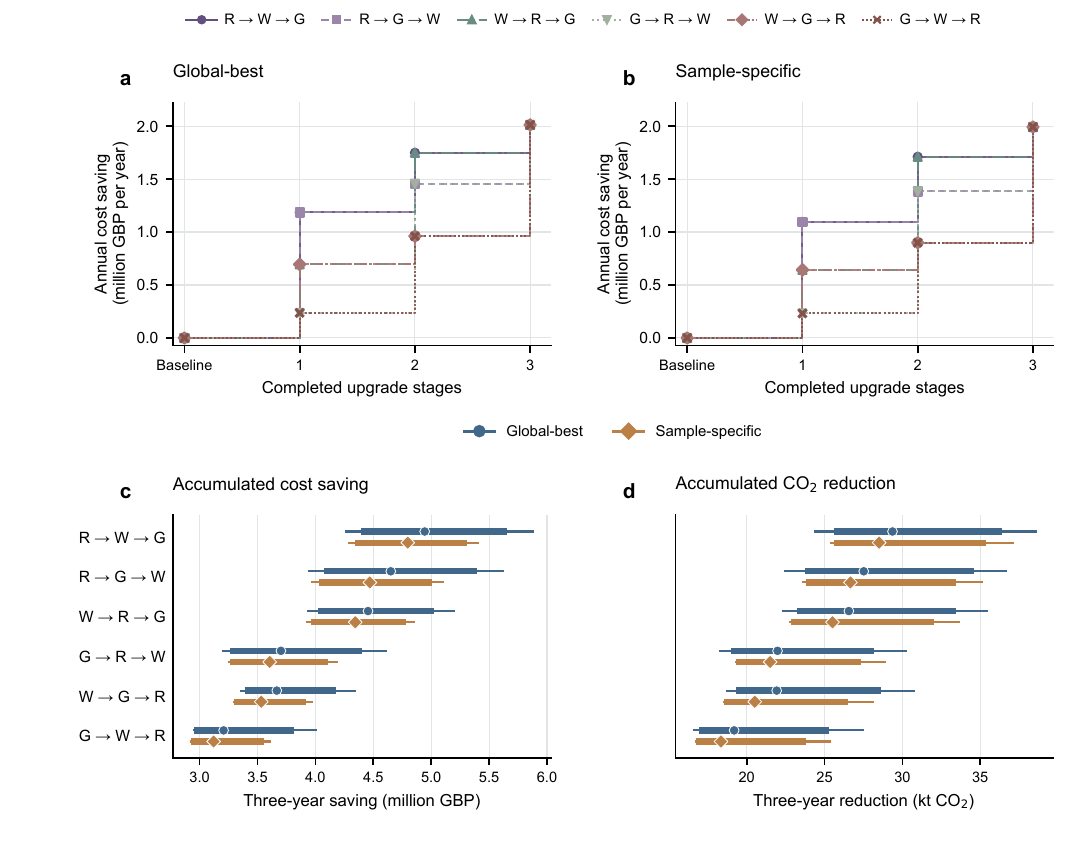}
\caption{Identical completed packages can yield different accumulated benefits. All six orders use the same 5,382 three-measure dwellings. \textbf{a,b}, Annual cost point benefits after each added measure, for global-best and the primary selector. Lines connect discrete stages and coincide at completion. \textbf{c,d}, Three-year accumulated cost savings and CO$_2$ reductions, assigning one year to each state. W denotes wall, G window glazing and R roof. Symbols show point totals; thick and thin lines show central 90\% and 95\% scenario ranges after summation. Orders follow global-best's cost point ranking. Pairwise contrasts are calculated within shared realisations, separately from these marginal ranges.}
\label{fig:scenario-staging}
\end{figure}

The next comparison holds the candidate population fixed and changes only the order of the measures. Among 8,626 dwellings needing all three measures, 5,382 had common support for every state. For these dwellings, all six orders ended with the same projected annual package benefit under global-best: \pounds2.01 million and 12.04\,kt CO$_2$. The order changed the intermediate states and the time for which each benefit accrued (Fig.~\ref{fig:scenario-staging}).

With one year at each stage, roof--wall--window accumulated the largest projected benefit and window--wall--roof the smallest under both strategies. Under global-best, three-year cost savings were \pounds4.94 million against \pounds3.21 million, a difference of \pounds1.737 million [1.447, 1.838], or 54.1\% more. Emission reductions were 29.37 against 19.19\,kt CO$_2$, a difference of 10.17\,kt [8.67, 11.09], or 53.0\% more. The primary selector gave corresponding relative advantages of 53.7\% and 55.3\%.

Earlier delivery of the larger isolated response explained most of the difference. Within this matched cohort, isolated mean annual cost savings under global-best were \pounds220.11 for roofs, \pounds129.14 for walls and \pounds43.85 for windows. Time weighting contributed \pounds1.897 million to the largest-minus-smallest order contrast, while interactions in the intermediate pairs reduced it by \pounds0.161 million, or 8.47\%. After roofs, adding walls also gave a larger increment than adding windows, so roof--wall--window exceeded roof--window--wall by \pounds0.293 million [0.258, 0.319].

\subsection{Engine agreement was closer for mean benefits than individual ranges}
\label{sec:engine-results}

The register's improvement calculation provides a check on how far these projections can be trusted. Certificates with exactly one corresponding recommendation supplied 8,616 wall, 1,277 window and 513 roof candidates on common support. Under global-best, mean SAP changes were smaller than the engine's by 0.49, 0.57 and 0.33 points respectively (Table~\ref{tab:engine-validation}). The engine's mean changes on these certificates ranked the measures in the same order as the projections, from roof to wall to window, and the primary selector gave similar results.

\begin{table}[!htbp]
\centering\footnotesize
\setlength{\tabcolsep}{4pt}
\caption{Projected single-measure score changes against the register's improvement calculation on common support. Register is the mean potential minus current rating for certificates with exactly one corresponding Appendix~T recommendation; Projected is the mean constrained point change. Difference is the mean paired difference (projected minus register), with its 95\% building-group bootstrap confidence interval. Mean $|\mathrm{Diff.}|$ is the mean absolute paired difference; $|\mathrm{Diff.}|\leq5$ is the percentage within five score points. Differences are in score points (Equations~\eqref{eq:engine-difference} and \eqref{eq:engine-metrics}).}
\label{tab:engine-validation}
\begin{tabular}{@{}lllrrrrrr@{}}
  \toprule
  Target & Measure & Strategy & $n$ & Register & Projected & Difference & Mean $|\mathrm{Diff.}|$ & $|\mathrm{Diff.}|\leq5$ \\
  \midrule
  SAP & Wall & Global-best & 8,616 & 4.92 & 4.43 & \shortstack{$-$0.49\\[-1pt]{\scriptsize [$-$0.56, $-$0.42]}} & 1.84 & 95.1\% \\
   &  & Primary selector & 8,616 & 4.92 & 4.54 & \shortstack{$-$0.38\\[-1pt]{\scriptsize [$-$0.45, $-$0.31]}} & 1.89 & 95.2\% \\
   & Window & Global-best & 1,277 & 2.70 & 2.14 & \shortstack{$-$0.57\\[-1pt]{\scriptsize [$-$0.67, $-$0.47]}} & 0.93 & 100.0\% \\
   &  & Primary selector & 1,277 & 2.70 & 2.13 & \shortstack{$-$0.58\\[-1pt]{\scriptsize [$-$0.68, $-$0.48]}} & 0.95 & 99.9\% \\
   & Roof & Global-best & 513 & 8.92 & 8.59 & \shortstack{$-$0.33\\[-1pt]{\scriptsize [$-$0.55, $-$0.08]}} & 1.95 & 95.1\% \\
   &  & Primary selector & 513 & 8.92 & 8.58 & \shortstack{$-$0.34\\[-1pt]{\scriptsize [$-$0.60, $-$0.07]}} & 2.11 & 94.3\% \\
  \midrule
  EI & Wall & Global-best & 8,616 & 6.92 & 5.89 & \shortstack{$-$1.03\\[-1pt]{\scriptsize [$-$1.12, $-$0.94]}} & 2.59 & 87.6\% \\
   &  & Primary selector & 8,616 & 6.92 & 5.94 & \shortstack{$-$0.98\\[-1pt]{\scriptsize [$-$1.08, $-$0.88]}} & 2.67 & 86.9\% \\
   & Window & Global-best & 1,277 & 3.89 & 2.99 & \shortstack{$-$0.90\\[-1pt]{\scriptsize [$-$1.06, $-$0.74]}} & 1.47 & 98.6\% \\
   &  & Primary selector & 1,277 & 3.89 & 2.95 & \shortstack{$-$0.94\\[-1pt]{\scriptsize [$-$1.11, $-$0.77]}} & 1.56 & 98.6\% \\
   & Roof & Global-best & 513 & 12.36 & 11.31 & \shortstack{$-$1.05\\[-1pt]{\scriptsize [$-$1.36, $-$0.71]}} & 2.66 & 86.7\% \\
   &  & Primary selector & 513 & 12.36 & 11.38 & \shortstack{$-$0.98\\[-1pt]{\scriptsize [$-$1.40, $-$0.56]}} & 3.06 & 79.5\% \\
  \bottomrule
\end{tabular}

\end{table}

Agreement across individual dwellings was weaker than agreement in mean changes, particularly for walls and windows: under global-best, Spearman correlations between projected and engine SAP changes were 0.48 for walls, 0.31 for windows and 0.85 for roofs. The central 90\% scenario ranges contained the engine's SAP change for only 22.9--45.3\% of matched certificates across measures and strategies (Supplementary Information).

\section{Discussion}
\label{sec:discussion}

The projections distinguish three features of a retrofit programme: how much each candidate dwelling responds, how many dwellings can contribute, and when benefits arrive. Westminster's housing stock and the delivery schedule explain why these features set different priorities. How far the estimates can be relied on depends on whether a programme uses a current score, the benefit at one dwelling or a portfolio total.

\subsection{Response size and applicability set different priorities}

Roof and wall rankings differ because portfolio benefit combines a mean response with the size of its candidate population. Westminster's stock is dominated by flats, and most dwellings have premises above, restricting direct roof upgrades (Fig.~\ref{fig:study-area}b). Roof candidates have larger projected responses, while walls reach many more dwellings. This distinction allows a programme to separate measures with large benefits for their candidates from measures contributing most across the available portfolio.

The recorded deficits offer a plausible explanation for the larger roof response. In the matched three-measure cohort, 93.05\% of dwellings had roof descriptions indicating no or zero-thickness insulation, whereas window edits included completing partly improved glazing. The engine's mean single-measure changes ranked the three measures in the same order as the projections, from roof to wall to window.

Projecting through reconstructed ratings keeps the comparison within the certificate assessment convention. Urban retrofit models such as the Dublin study evaluate options against modelled energy outcomes directly \citep{ali2020datadriven}; this study instead converts SAP and EI changes into standardised cost and operational emissions. The projected wall reduction of 36.34\,kt CO$_2$ per year is 15.8\% of Westminster's 2024 domestic emissions of 229.99\,kt \citep{desnz2026local}, a comparison of scale between a rating-equivalent projection and a territorial inventory.

Rating bands offer a further programme reporting unit. For example, 14,480 of 23,534 candidates with projected baseline SAP in band D were projected to reach C or above after a wall edit (Supplementary Information). Only 23 supported wall candidates lay below band E, so these data give little evidence about programmes targeting the private-rented minimum standard \citep{desnz2026mees}.

\subsection{Earlier benefits distinguish orders with an identical endpoint}

The sequence result concerns the timing of delivery. Under the same three one-year stages, roof--wall--window accumulated 54.1\% more projected cost saving than window--wall--roof, although both finished at the same annual benefit. Giving the larger response more time explained most of that contrast. Interactions reduced the advantage without reversing it, and the larger increment from walls after a roof edit distinguished the two roof-first orders. The comparison therefore addresses what a programme delivers while implementation is under way, as well as what it delivers when complete.

After completion, the accumulated absolute gap remains while its relative contribution declines with a longer horizon. Stock-trajectory optimisation can incorporate measures and timing into wider objectives \citep{rit2024shape}; investment optimisation would add installation costs, financing and practical capacity constraints.

For emissions, the sign of non-additivity depended on the reported scale. Under global-best, the combined EI-score gain exceeded the sum of isolated gains by 0.378 points per dwelling, yet the converted emission reduction was 0.474\,kt CO$_2$ per year below the isolated-benefit sum. Relative to a linear reference, the curvature of the emission conversion accounts for this sign reversal (Supplementary Information). Coarse descriptions lack the geometry and thermal detail needed to identify physical synergy or diminishing returns, or the interactions between buildings and urban context examined in hybrid modelling \citep{nutkiewicz2021urbancontext}.

\subsection{Reliability depends on the quantity being used}

Three different questions arise along the prediction-to-benefit calculation. Current-score evaluation asks whether intervals contain registered outcomes. The primary selector's 90\% intervals achieved 92.0\% SAP and 91.0\% EI coverage, while some observable subgroups fell below nominal. This is compatible with marginal rather than dwelling-conditional coverage \citep{lei2018distribution,barber2021limits}. Width matters alongside coverage: wider intervals can cover more scores while providing less precision, as the full six-family comparison shows \citep{gneiting2007proper}.

The next question is the uncertainty of a before/after difference. The primary selector's mean converted 90\% current-score span was \pounds362.34 per year, compared with global-best's mean candidate savings of about \pounds49--203. This span measures uncertainty in a single state, whereas a dwelling's benefit range also depends on how the errors of its baseline and edited states move together. Building-energy studies can propagate uncertainty in physical parameters or construct intervals directly for loads \citep{guo2025uncertainty,massidda2023probabilistic}; here shared rank supplies a dependence assumption for uncertain reconstructed scores.

The engine comparison checks the paired projections against an external calculation. Under global-best, mean SAP changes fell short by 0.3--0.6 points, but individual ranges contained substantially fewer engine changes than their nominal percentages. The weaker rank agreement for walls and windows suggests that mean agreement can conceal missing dwelling-specific variation. Unpublished heat-loss wall area is one plausible contributor, approximated only indirectly by the recorded descriptors. Differences between Appendix~T interventions and coarse edits are another plausible contributor.

The third question is how errors move together across dwellings when benefits are summed. Sharing a rank across the whole portfolio prevents reliance on independent score errors averaging out. No available reference tests this assumption. The projections therefore support comparisons between measures and orders more firmly than the benefit at one dwelling or the range of a portfolio total.

\subsection{Both predictors gave similar planning comparisons}

Global-best supplied the accuracy reference; sample-specific selection tested whether local error information should change the family used for a dwelling. The latter enlarged supported reach but raised SAP MAE and showed no demonstrable EI improvement. Nevertheless, roofs led per candidate and walls in total under both strategies, and all six staged orders retained their ranking. Only the close roof and window emission totals changed places. The principal planning comparisons were therefore similar across these two choices of predictor.

The accuracy and fitting-resource comparison also matters when choosing the prediction reference (Fig.~\ref{fig:accuracy-cost}). CatBoost accepts all 13 string fields as native categorical variables, whereas Ridge, Random Forest, LightGBM and XGBoost use one-hot encoding. Its categorical-feature processing is a plausible contributor to its accuracy with these descriptions \citep{prokhorenkova2018catboost}. Multimodal learns text representations and categorical embeddings but did not surpass the leading tree models. CatBoost required 1,085\,s across both targets against LightGBM's 25\,s for MAE reductions of 0.087 SAP and 0.056 EI points. LightGBM is therefore a useful development reference where fitting resources are limited.

The selection results distinguish model complementarity from the ability to exploit it for a new dwelling (Fig.~\ref{fig:selection-tradeoff}). The outcome-informed oracle observes outcomes and includes operationally ineligible candidates, so its advantage can reflect unpredictable error variation as well as learnable specialisation. A deployable selector has only the new description and development evidence from which to identify relative error. Noisy local risk estimation is a plausible explanation: the selected neighbourhoods used 100, 200 or 500 neighbours, and shrinking towards global error does not ensure that local model errors are ordered correctly. Selector tuning and neighbourhood development also optimised serious-error and interval-score objectives rather than the paired MAE used for evaluation. Prediction accuracy and downstream objectives need not coincide \citep{fan2026decision}.

Broader support is a separate outcome from greater accuracy: an alternative family can return a supported prediction when CatBoost cannot. Requiring common support in the scenario comparison (Fig.~\ref{fig:scenario-applicability}) holds the comparison population fixed and lets the two strategies test sensitivity to predictor choice. Future work could distinguish switches forced by CatBoost's ineligibility from choices made while it remained eligible, and examine whether relative-error calibration improves routing.

\subsection{Limitations and scope}

The empirical evidence concerns Westminster, detailed certificate descriptors and one fitted interval system. Historical residuals mix model error, unrecorded attributes and certificate measurement variation \citep{crawley2019measurement,hardy2019errors}; they do not measure paired errors before and after installation. The engine comparison uses selected single-recommendation certificates, including only 513 roof cases.

Scenario magnitudes come from learned associations. Description vocabulary and neighbourhood support are consequential, particularly when entire strings are categorical features. Corrections as large as 38.57 EI points show that small average adjustments can hide substantial individual conflicts with the assumed physical direction. Coarse edits do not determine construction details, heritage consent or feasibility, and shared walls and roofs require coordination. Workmanship, occupancy response and altered service demand lie outside the fixed-service scenario. Monetary quantities are standardised annual energy-cost savings without installation costs.

\section{Conclusion}
\label{sec:conclusion}

Model-based fabric retrofit projections give different priorities when viewed per candidate dwelling, across a housing portfolio and over time. In Westminster, roof insulation delivered the largest projected mean annual cost saving per candidate under global-best, about \pounds203, while wall insulation had the largest total, \pounds6.09 million per year [5.69, 6.72] on common support, through wider applicability. On 5,382 supported dwellings needing all three measures, roof--wall--window accumulated 54.1\% more projected cost saving than window--wall--roof over three one-year stages. All orders ended with the same annual package benefit; their difference was when savings arrived.

Global-best remained the accuracy reference, while sample-specific selection extended support and gave similar principal scenario rankings. After independent post-selection calibration, the current-score intervals of both strategies reached nominal marginal coverage. Comparing benefits rested on an additional shared-rank assumption, applied before paired conversion and aggregation. The register's engine calculations were closer to projected mean changes than to individual variation, and the scenario ranges did not attain their nominal percentages against that reference. For a retrofit programme, the projections are therefore better suited to ranking measures and orders than to estimating the benefit at one dwelling. Paired pre- and post-retrofit assessments would allow actual savings and the dependence between baseline and edited errors to be evaluated. Evidence on how errors move together across dwellings would extend this evaluation to portfolio totals.

\section*{Acknowledgements}
The authors would like to thank the funding support from Westminster City Council and King's College London.
\setlength{\bibsep}{3pt}
\bibliographystyle{unsrtnat}
\bibliography{references}
\end{document}